\documentclass[preprint,12pt]{elsarticle}
\usepackage{latexsym, graphicx, epsfig, amsmath, amssymb, amsfonts}
\usepackage{amsmath}
\usepackage{amssymb}
\usepackage{graphicx}
\usepackage{subfig}
\usepackage{tabu}
\usepackage{adjustbox}
\usepackage{bbold}
\usepackage{bm}
\usepackage{textcomp}
\usepackage[thinc]{esdiff}
\usepackage[dvipsnames]{xcolor}
\usepackage{comment}
\usepackage{textgreek}
\usepackage{caption}
\usepackage{lineno}
\usepackage{multirow}
\usepackage{adjustbox}
\usepackage{amsmath}

\usepackage{siunitx}
\usepackage{xcolor}
\usepackage{booktabs}
\usepackage{caption}
\usepackage{float}
\usepackage{color,soul}

\journal{The Journal}
\begin{document}
\begin{frontmatter}

\title{Identifying Constitutive Parameters for Complex Hyperelastic Materials using Physics-Informed Neural Networks}

\author{Siyuan Song\textsuperscript{a}}
\author{Hanxun Jin\textsuperscript{a,b,}\corref{cor1}}
\cortext[cor1]{Corresponding authors. E-mail:  hanxun\_jin@alumni.brown.edu (H.J.)}

\address[1]{School of Engineering, Brown University, Providence, RI 02912}
\address[2]{Present Address: Division of Engineering and Applied Science, California Institute of Technology, Pasadena, CA 91125}

\begin{abstract}
Identifying constitutive parameters in engineering and biological materials, particularly those with intricate geometries and mechanical behaviors, remains a longstanding challenge. The recent advent of Physics-Informed Neural Networks (PINNs) offers promising solutions, but current frameworks are often limited to basic constitutive laws and encounter practical constraints when combined with experimental data. In this paper, we introduce a robust PINN-based framework designed to identify material parameters for soft materials, specifically those exhibiting complex constitutive behaviors, under large deformation in plane stress conditions. Distinctively, our model emphasizes training PINNs with multi-modal synthetic experimental datasets consisting of full-field deformation and loading history, ensuring algorithm robustness even with noisy data. Our results reveal that the PINNs framework can accurately identify constitutive parameters of the incompressible Arruda-Boyce model for samples with intricate geometries, maintaining an error below 5\%, even with an experimental noise level of 5\%. We believe our framework provides a robust modulus identification approach for complex solids, especially for those with geometrical and constitutive complexity.
\end{abstract}

\end{frontmatter}


\newpage
\section{Introduction}
The emerging demands to characterize biological materials and design advanced bio-inspired metamaterials necessitate precisely identifying their material properties, such as Young’s modulus and shear modulus. Traditionally, the modulus is identified through mechanical tests \cite{davis2004tensile}, such as uniaxial tensile experiments performed on specially prepared samples that comply with testing protocols. This approach involves collecting stress-strain or pressure-volume relationships \cite{jin2021ruga} and presuming a phenomenological constitutive model for the material, such as the Neo-Hookean (NH) model \cite{treloar1948stresses}. Subsequently, the material modulus can be identified by fitting the experimental data with the presumed model. As the complexity of the constitutive model increases (e.g., Fung-type model \cite{fung1979pseudoelasticity}, Holzapfel-Gasser-Ogden (HGO) model \cite{holzapfel2000new}), a variety of testing methodologies, including shear and biaxial testing, become necessary to accurately identify the constitutive parameters \cite{sacks2000biaxial,sacks2003multiaxial}. Nevertheless, this traditional framework requires a strict sample preparation process. For example, samples should have a carefully designed dog-bone shape to allow parameter identification under uniaxial testing. However, shaping certain biological tissues like cruciate ligaments into such specific geometry proves difficult \cite{luetkemeyer2021constitutive}. Furthermore, the advent of novel materials like mechanical metamaterials \cite{jin2023mechanicalmeta,bertoldi2017flexible}, which contain intentionally designed defects, makes conventional methodologies, like finite element model update methods or virtual field methods \cite{avril2008overview}, insufficient for identifying their homogenized material parameters. This inadequacy primarily arises from the inherent difficulties associated with measuring internal displacement fields in such complex structures. Consequently, identifying the material parameters for samples with intricate geometries comprising irregular boundaries and internal defects continues to have challenges. 

Recent advances in deep learning have seen a surge in the application of end-to-end material properties identifications \cite{gu2018bioinspired, jin2022dynamic, ni2021deep,bock2019review,yang2021end,jin2022big,guo2021artificial,liu2020machine,yang2020prediction,jin2023mechanical}. In most cases, these methods require a sufficient training dataset, sourced primarily from simulations. However, this approach often requires significant computations, particularly when working with high-dimensional parameter spaces \cite{jin2023review}. While traditional data-driven methods embed fundamental physics, such as governing equations, into simulations, there is an increasing demand for more effective methodologies. The focus has shifted towards methods that directly encode these physical laws into deep learning algorithms, enhancing the efficiency of material modulus identification. To this end, Physics-Informed Neural Networks (PINNs) \cite{raissi2019physics, karniadakis2021physics} have been developed to explicitly encode the underlying physical laws, i.e., partial differential equations (PDEs), into the neural networks (NNs). This integration of prior physical law knowledge into a deep learning framework can notably minimize the required size of the training dataset without compromising prediction accuracy. Since their inception, PINNs have demonstrated substantial utility across a broad spectrum of engineering disciplines, including solid mechanics \cite{haghighat2021physics, zhang2020physics, zhang2022analyses, li2021physics, bastek2023physics, chen2023physics, henkes2022physics,niu2023modeling}, fluid mechanics \cite{cai2021physics, raissi2020hidden, cai2021artificial, jin2021nsfnets}, and biomaterials \cite{kamali2023elasticity, yin2021non}. As an example of the material modulus identifications, Zhang et al. \cite{zhang2020physics, zhang2022analyses} demonstrated that PINNs could efficiently identify the distribution of inhomogeneous material and geometry under the plane strain condition, bypassing the need for computation-intensive data-driven processes. These PINNs frameworks, specifically tailored for continuum solid mechanics, hold promising potential to address the identified challenges in modulus identification of solids with intricate constitutive models. As such, they represent an exciting frontier in material properties identification. However, it is important to note that the current PINNs frameworks still have their limitations, with their capacity largely restricted to identifying parameters from simple constitutive laws such as linear elasticity \cite{kamali2023elasticity,anton2022physics, chen2023physics} and NH solids \cite{zhang2020physics}. Recent endeavors by Hamel et al. \cite{hamel2023calibrating} seek to enhance the original PINNs framework to calibrate constitutive parameters for complex hyperelastic models using full-field measurements. However, the adoption of a computational grid to integrate the PDE's weak form poses challenges when combined with experimental full-field measurement such as Digital Image Correlation (DIC). Therefore, a robust and applicable PINN-based algorithm for identifying constitutive parameters of complex hyperelastic solids is still lacking.

Herein, we introduce a PINN-based deep learning framework, which can accurately identify material parameters for complex constitutive behaviors beyond NH from multi-modal synthetic experimental data. Compared to the original framework \cite{zhang2022analyses}, we integrated the complete deformation and loading history into the PINNs under the plane stress condition, which is more practical to the experimental setup. To accommodate inhomogeneous deformation, particularly when substantial internal defects are present, we coupled the full-field displacements measured from DIC \cite{chu1985applications} as additional boundary conditions (BCs) for training. These full-field displacement measurements provide valuable local deformation data induced by internal inhomogeneities. Therefore, by combining the full-field displacement field into the PINNs framework, we can effectively predict material characteristics even in scenarios where conventional methods prove insufficient due to the complexity of biological samples. In \textbf{Section 2}, we will briefly review the fundamentals of PINNs for continuum solid mechanics and present our detailed algorithm. In \textbf{Section 3}, we will first verify our framework by solving a forward problem and compare the results with those obtained from finite element methods (FEM). Subsequently, we will employ the framework to identify the constitutive parameters for Arruda-Boyce (AB) constitutive laws using prototype problems: rectangular samples with circular inhomogeneities inside. We will also discuss the quality and noise effects of DIC data on the overall accuracy and efficiency of prediction. We anticipate that our framework will provide a robust modulus identification approach for soft materials with complex constitutive behaviors and geometries using direct experimental measurements.

\section{PINNs framework for parameter identification}

\begin{figure}[!ht]
    \centering
    \includegraphics[width=1.0\textwidth]{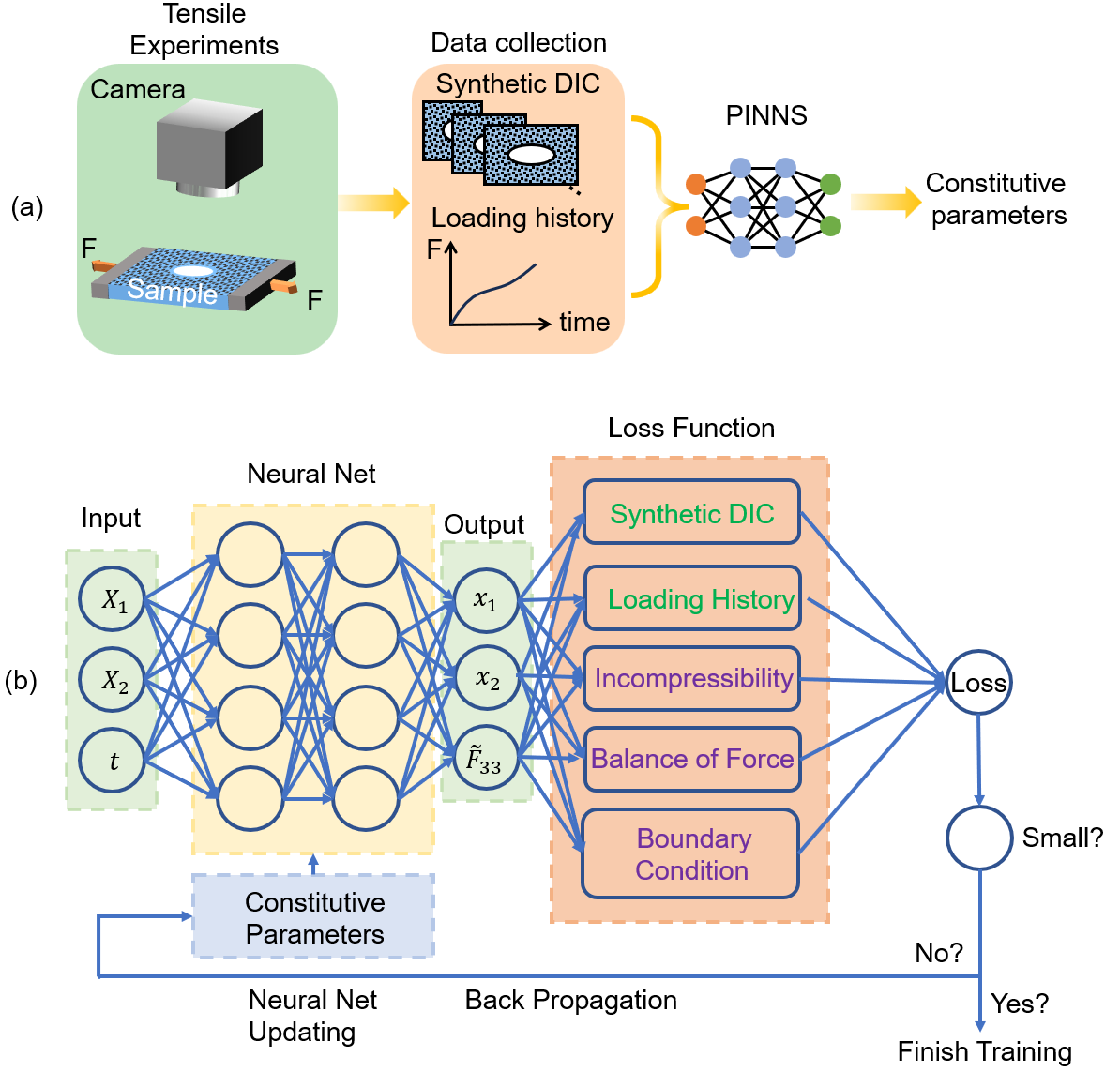}
    \caption{(a) The schematics of experimental data collection for PINNs. A sample with internal geometry defects is subjected to displacement BCs. A camera is used to record the DIC dataset and a load cell is used to record the loading history. Then, PINNs incorporate these two experimental datasets during training to identify the constitutive parameters. (b) The architecture of PINNs for samples under the plane stress condition.}
    \label{fig:Figure-1}
\end{figure}

The methodology for data collection and processing for our algorithm is illustrated in \textbf{Fig.1(a)}. A sample with internal geometric anomalies and unidentified material parameters is subjected to uniaxial tension under the plane stress condition. The sample's top surface is speckled and subsequently captured with a digital camera. DIC analysis \cite{chu1985applications} can then be employed to obtain the full-field displacements. Simultaneously, a load cell records the loading under deformation. To the end, a multi-modal experimental dataset can be collected, containing both the full-field displacement field and loading history. This dataset subsequently serves as additional BCs for the PINNs training, aiding in identifying constitutive parameters. In \textbf{Section 2.1}, we first introduce the PINNs framework, tailored for continuum solids with hyperelastic constitutive behaviors under large deformation. Then, we encompass the setup for a prototype problem in \textbf{Section 2.2}, utilized to validate our framework via both forward surrogate modeling and inverse parameter identification. In this paper, we used FEM to generate the synthetic multi-modal experimental data to validate our framework.

\subsection{PINNs for continuum solid mechanics}
Here, we briefly review PINNs for continuum solid mechanics and introduce our modifications for complex hyperelastic solids beyond simple NH materials under the quasi-static plane stress deformation. The readers can refer to recent research papers \cite{raissi2019physics, raissi2020hidden, zhang2022analyses} and reviews \cite{jin2023review,cai2021physics} for detailed descriptions of PINNs. The architecture of PINNs is depicted in \textbf{Fig.1(b)}. For convenience, we denote directions 1, 2, and 3 as the loading direction, transverse direction, and out-of-plane direction, respectively. We define the unknown constitutive parameters as $\boldsymbol\theta_{\text{const}}$. There are in total four key steps to identify $\boldsymbol\theta_{\text{const}}$ in the PINNs framework. 

First, we employ a neural network (NN) (Net $\mathcal{U}$) to approximate the solutions of mapping the coordinates in the undeformed configuration $\textbf{X} = (X_1, X_2)$ to the coordinates in the deformed configuration $\tilde{\textbf{x}} = (\tilde{x}_1,\tilde{x}_2)$ and out-of-plane deformation gradient $\tilde{F}_{33}$ as,
\begin{equation}
(\tilde{\textbf{x}},\tilde{F}_{33}) = (\tilde{x}_1,\tilde{x}_2,\tilde{F}_{33}) = \mathcal{NN_U}(\textbf{X},t;\boldsymbol\theta_{NN},\boldsymbol\theta_{\text{const}}).
\end{equation}
Here, we use the tilde sign to represent an approximate quantity from the NN. $\boldsymbol\theta_{NN}$ are trainable parameters of the NN (weights and biases). The quasi-static stretch ratio $\Lambda$ is proportional to the unitless time measure $t$. The displacement at position $\textbf{X}$ can be calculated as,
\begin{equation}
\tilde{u}(\textbf{X};t)=\tilde{\textbf{x}}-\textbf{X}. 
\end{equation}

Second, we integrate continuum mechanics into the PINNs by deriving relevant quantities from the NN outputs. During the training process, the partial derivatives are processed by automatic differentiation \cite{baydin2018automatic}. Therefore, the deformation gradient $\tilde{\textbf{F}}(\textbf{X};t)$ and the first Piola-Kirchhoff stress $\tilde{\textbf{P}}(\textbf{X};t)$ of a generalized hyperelastic solid can be calculated as,

\begin{equation}
\tilde{\textbf{F}}(\textbf{X};t)=\frac{\partial{\tilde{\textbf{x}}}}{\partial{\textbf{X}}}
\end{equation}
\begin{equation}
\tilde{\textbf{P}}(\textbf{X};t)=-\tilde{p}(\textbf{X};t)\tilde{\textbf{F}}^{-T}(\textbf{X};t)+2\tilde{\textbf{F}}(\textbf{X};t)\frac{\partial\Phi(\textbf{X};t)}{\partial\tilde{\textbf{C}}(\textbf{X};t)}
\end{equation}
where $\tilde{p}(\textbf{X};t)$ is the hydrostatic pressure, which can be solved from BCs in the plane stress condition. $\tilde{\Phi}(\textbf{X};t)$ is the strain energy density, and $\tilde{\textbf{C}}(\textbf{X};t) = \tilde{\textbf{F}}^T(\textbf{X};t)\tilde{\textbf{F}}(\textbf{X};t)$ is the right Cauchy-Green deformation tensor. Then, the static equilibrium condition for a continuum solid body and material incompressibility require,
\begin{equation}
\mathrm{Div} \tilde{\textbf{P}}(\textbf{X};t)\approx0;
\end{equation}
\begin{equation}
\mathrm{Det} \tilde{\textbf{F}}(\textbf{X};t)\approx0.
\end{equation}
Next, we consider the prescribed Dirichlet (displacement) and Neumann (traction) BCs as,
\begin{equation}
\tilde{\textbf{u}}(\textbf{X};t)\approx\bar{\textbf{u}}(\textbf{X};t)
\end{equation}
\begin{equation}
\tilde{\textbf{P}}(\textbf{X};t)\textbf{N}(\textbf{X})\approx\bar{\textbf{T}}(\textbf{X};t)
\end{equation}
where $\textbf{N}(\textbf{X})$ is the normal unit vector on the boundary at position $\textbf{X}$. $\bar{\textbf{u}}(\textbf{X};t)$ and $\bar{\textbf{T}}(\textbf{X};t)$ are the prescribed displacement and traction on the corresponding boundary at time $t$. Furthermore, we apply the discrete displacement dataset, $\textbf{u}^*(\textbf{X}_{\text{DIC}};t)$, measured from speckle images through DIC at position $\textbf{X}_{\text{DIC}}$ as additional BCs, which we expect that they match the PDE solutions, $\tilde{\textbf{u}}(\textbf{X}_{\text{DIC}};t)$. Furthermore, we expect the total traction on the boundary that is subjected to the uniaxial tension, $\sum_{\partial\Omega_t}\bar{\textbf{T}}(\textbf{X};t)$, to match the loading history measured from experimental data, $\textbf{T}^*(\textbf{X};t)$. For the large deformation case we considered here, we found the extraction of the force from a specific boundary leads to a bias in the evaluation of the PDEs in the whole domain, where the sample can artificially develop necking during PINNs training. To eliminate this numerical problem, we adopted a Monte Carlo based domain integral to calculate the loading force. Specifically, the loading per unit length in $X_1$ direction is calculated as,
\begin{equation}
T_1(t)\approx\frac{1}{\Omega_p}\sum_{\Omega_p}\tilde{T}_1(X_1;t)
\end{equation}
where $\Omega_p$ is a set of the points randomly sampled from the domain away from the boundary and internal inhomogeneities.

In the third step, we can formulate the loss function at each training step as,
\begin{equation}
    \mathcal{L}=\alpha_1 \mathcal{L}^{\text{PDE}}+\alpha_2 \mathcal{L}^{\text{inc}}+\alpha_3 \mathcal{L}^{\text{BC}}+\alpha_4 \mathcal{L}^{\text{Loading}}+\alpha_5 \mathcal{L}^{\text{DIC}}
\end{equation}
where $\mathcal{L}^{\text{PDE}}$, $L^{\text{inc}}$, $L^{\text{BC}}$, $L^{\text{Loading}}$, and $L^{\text{DIC}}$ are the loss functions of the PDEs, incompressibility, boundary conditions, loading history, and displacements measured from DIC, respectively. $\alpha$'s are their corresponding weights. Their loss can be defined as,
\begin{equation}
    \mathcal{L}^{\text{PDE}}=\dfrac{1}{N_{\text{PDE}}}\sum_{i=1}^{N_{\text{PDE}}} |\mathrm{Div} \tilde{\textbf{P}}(\textbf{X};t)|^2
\end{equation}
\begin{equation}
    \mathcal{L}^{inc}=\dfrac{1}{N_{\text{PDE}}}\sum_{i=1}^{N_{\text{PDE}}} |\mathrm{Det} \tilde{\textbf{F}}(\textbf{X};t)|^2
\end{equation}
\begin{equation}
    \mathcal{L}^{\text{BC}}=\dfrac{1}{N_{\text{BCU}}}\sum_{i=1}^{N_{\text{BCU}}} |\tilde{\textbf{u}}(\textbf{X};t)-\bar{\textbf{u}}(\textbf{X};t)|^2 + \dfrac{1}{N_{\text{BCP}}}\sum_{i=1}^{N_{\text{BCP}}} |\tilde{\textbf{P}}(\textbf{X};t)\textbf{N}(\textbf{X})-\bar{\textbf{T}}(\textbf{X};t)|^2
\end{equation}
\begin{equation}
    \mathcal{L}^{\text{Loading}}=\frac{1}{N_{\text{Loading}}} \sum_{i=1}^{N_{\text{Loading}}}(T_1(t)-T_1^*(t))^2
\end{equation}
\begin{equation}
    \mathcal{L}^{\text{DIC}}=\dfrac{1}{N_{\text{DIC}}}\sum_{i=1}^{N_{\text{DIC}}} |\tilde{\textbf{u}}(\textbf{X}_{\text{DIC}};t)-\textbf{u}^*(\textbf{X}_{\text{DIC}};t)|^2.
\end{equation}
Here, $N_{\text{PDE}}$, $N_{\text{BCU}}$, $N_{\text{BCP}}$, $N_{\text{Loading}}$, and $N_{\text{DIC}}$ are the number of sampling points inside the PDE domain, on the boundaries subjected to displacement BCs, on the boundaries subjected to traction BCs, number of loading measurement, and number of full-field displacement measurement.   

In the last step, we can minimize the total loss through NN and constitutive parameter optimization using the following notation,
\begin{equation}
    \hat{\boldsymbol{\theta}}_{NN}, \hat{\boldsymbol{\theta}}_{\text{const}} =\underset{\boldsymbol{\theta}_{NN},\boldsymbol{\theta}_{\text{const}}}{\arg\min}\mathcal{L}(\boldsymbol{\theta}_{NN}, \boldsymbol{\theta}_{\text{const}})
\end{equation}
where the hat symbol represents the optimized parameters.

\subsection{Setup of the prototype problem}
To validate our proposed PINNs framework in parameter identification, we examine a prototype model: a rectangle featuring an internal hole subjected to displacement BCs under the plane stress condition. As shown in \textbf{Fig.2(a)}, the length and width of the sample are 2$a$ and 2$b$, and the coordinates of four corners are $(\pm a,\pm b)$. Some circular inhomogeneities (i.e., holes or different materials) exist within the domain with radius $r$. We consider the medium exhibiting an incompressible hyperelastic constitutive behavior with unknown parameters. Here, as a proof of concept, we select the two-parameter AB 8-chain model \cite{arruda1993three} for the medium. The AB model, developed based on statistical mechanics, is a link between microstructural polymer physics and their resulting macroscopic deformations. Given its broad adaptability across various elastomeric materials, it is a prototype model in polymer mechanics, especially when the strain-stiffening phenomenon is distinct. The strain energy density function of the AB model is \cite{arruda1993three},
\begin{equation}
\Phi=Nk_BT\sqrt{n}[\beta\lambda_{\text{chain}}-\sqrt{n}\mathrm{ln}(\frac{\mathrm{sinh}\beta}{\beta})]
\end{equation}
where $N$ is the number of chains in polymer network, $n$ is the number of chain segments, $k_B$ is the Boltzmann constant, and $T$ is the temperature. Here, $\lambda_{\text{chain}}=\sqrt{\frac{I_1}{3}}$ with $I_1$ the first invariant of left Cauchy-Green tensor. $\beta=\mathcal{L}^{-1}(\frac{\lambda_{\text{chain}}}{\sqrt{n}})$ with $\mathcal{L}^{-1}(\bullet)$ the inverse Langevin function. If we consider the first five terms of the inverse Langevin function, the strain energy density can be approximated as \cite{systemes2007abaqus},
\begin{equation}
\Phi=\mu[\frac{1}{2}(I_1-3) + \frac{1}{20\lambda_m^2}(I_1^2-9) + \frac{11}{1050\lambda_m^4}(I_1^3-27) + \frac{19}{7000\lambda_m^6}(I_1^4-81) + \frac{519}{67375\lambda_m^8}(I_1^5-243)].
\end{equation}
We define the effective shear modulus as,
\begin{equation}
\mu^{*} = \mu(1.0 + \frac{1}{5\lambda_m^2}I_1 + \frac{11}{175\lambda_m^4}I_1^2 + \frac{19}{875\lambda_m^6}I_1^3 + \frac{519}{67375\lambda_m^8}I_1^4).
\end{equation}
Hence, the constitutive parameters for AB model consist of $\boldsymbol\theta_{\text{const}}=(\mu,\lambda_m)$.

\subsection{Training of PINNs}
We implemented and customized the PINNs framework on the open-source Python library, Deepxde 1.7.0, developed by Lu et al. \cite{lu2021deepxde}. The PINNs architecture has a total of 3 hidden layers with 30 neurons in each layer. In addition, we set \textit{Tanh} as the activation function and \textit{Glorot uniform} as the weight initializer. Through this paper, we set Adam \cite{kingma2014adam} as the optimization algorithm with a learning rate of 0.001. Other higher-order optimization algorithms like L-BFGS can be also employed \cite{liu1989limited} to further improve training efficiency. The PINNs were trained on a laptop with a 14-core i7-12700H CPU and a RTX3060 GPU. Leveraging GPU acceleration, we achieved a training speed of approximately 50 epochs per second.

\subsection{Finite Element Analysis}
We performed the FEM using Abaqus 2021 Static package \cite{systemes2007abaqus} to verify the forward problems and also generate the synthetic DIC and loading history dataset for the inverse problem. The medium was modeled using the incompressible AB constitutive law. The left boundary was clamped, while the right boundary had a displacement BC to deform the sample. The simulation domain was meshed with four-node plane stress elements (CPS4R). The linear mesh density is approximately 111 per unit length and mesh refinement near the internal defects was performed. Mesh dependence of the loading curve and the displacement field was examined, and all solutions reported are final converged solutions. The loading history was obtained by integrating the total force on the right boundary, and synthetic DIC data was extracted from the displacement field at the FEM nodes. We applied a Gaussian random noise with the desired level for the cases considering the DIC noise effects. 

\section{Results}
In this section, we first verify our PINNs algorithm by solving a forward problem and compare its results with those calculated from FEM. Then, we discuss its robustness in identifying the constitutive parameters using synthetic experimental data with and without noise. Building upon the modeling presented in \textbf{Section 2}, the outputs of the PINNs, $(f,g,\tilde{F}_{33})$, are subsequently subjected to a transformation resulting in $(x_1,x_2,\tilde{F}_{33})$, achieved through the following equations,
 \begin{equation}
     x_1=[\Lambda(X_1+b)-b]+(X_1+b)(X_1-b)f
 \end{equation}
 \begin{equation}
     x_2=X_2 + (X_1+b)(X_1-b)g
 \end{equation}
where $\Lambda$ is the stretch ratio, {$f$ and $g$ represent the variables that serve as intermediate outputs from the NNs}. The resulting $(x_1, x_2)$ inherently satisfies the displacement BCs \cite{lu2021deepxde}. This transformation can reduce the total number of loss functions during training and also ensure that the trivial solution, $f,g=0$, is close to the theoretical solution, leading to the reduction of the error in the power term $\frac{I_1^n}{\lambda_m^{2n}} (n=1,2,3,...)$. Furthermore, we consider the following unitless material and geometry parameters, $\lambda_m = 3$, $\mu = 1$, $a = 1$, $b = 0.5$, and the maximum stretch $\Lambda_{max} = 2.5$. In biomechanics literature, the strain hardening parameter $\lambda_m$ of the AB model is measured as $1.6$ for brain white tissues \cite{Ashrafi2017Arruda}, and $3.32 \pm 3.56$ for fibroadenoma \cite{o2009measurement}. The selected value $\lambda_m = 3$ is within the typical range for soft tissues. Through this paper, we normalized the loss functions by dividing them by the total loss at the initial step.

\subsection{Forward Problem}
The performance for material identification depends on the PINNs' capability to accurately predict the deformation field of the sample. To this end, in this forward problem, we investigate a rectangular domain featuring a centrally located hole with a radius of $r = 0.15$, as illustrated in \textbf{Fig.2(a)}. The parameters of PINNs, $\boldsymbol{\theta}_{NN}$, were optimized by minimizing the loss functions of incompressibility  $\mathcal{L}^{\text{inc}}$, equilibrium equations $\mathcal{L}^{\text{PDE}}$, displacement BCs $\mathcal{L}^{\text{BC1}}$ at $X_1 = \pm0.5$, and traction-free BCs $\mathcal{L}^{\text{BC2}}$ at $r = 0.15$ and $X_2 = \pm 0.5$. The associated weights for these loss functions were tuned to yield the best performance. Our training dataset comprised 20,000 randomly sampled points within the time-spatial domain, supplemented by 50 uniformly sampled points along the hole boundary and 200 uniformly sampled points along the free boundary, spanning across 21 evenly spaced time steps. We trained the PINNs for 1,000k epochs.

\begin{figure}[!ht]
    \centering
    \includegraphics[width=1.0\textwidth]{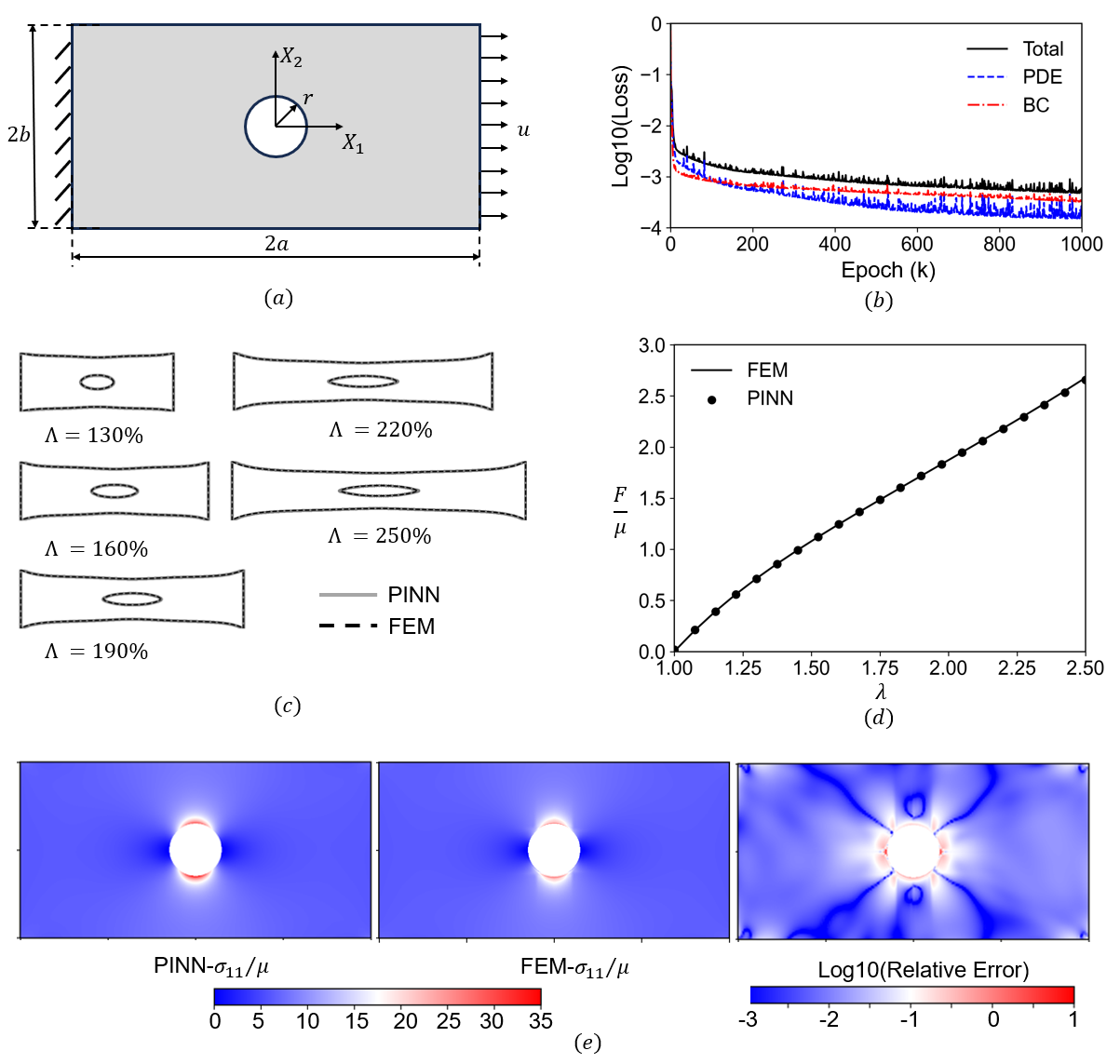}
    \caption{Forward problem of a rectangular sample with a central hole under the plane stress condition: (a) Schematics of the sample geometry and dimensions; (b) Loss as a function of training epochs. The solid line represents the total loss, the dotted line represents the loss of PDE, and the dashed line represents the loss of BCs; (c) Visual outlines of deformed configurations calculated from PINNs (solid lines) and FEM (dashed lines) under five different stretch ratios; (d) Loading as a function of stretch obtained from the PINNs (filled circles) and FEM (solid line); (e) The normalized Cauchy stress contour in the loading direction $\sigma_{11}$ of PINNs and FEM, and their logarithmic relative error. The contour is plotted in the undeformed configuration.}
    \label{fig:Figure-2}
\end{figure}

\textbf{Fig.2(b)} shows the losses rapidly reach $10^{-3}$ after 200k training epochs and converge to $10^{-3}-10^{-4}$ after 1,000K training epochs, which indicates our PINNs can optimize the NN parameters with incorporated physics efficiently. The deformation outlines under five different stretch ratios, as depicted in \textbf{Fig.2(c)}, showcase a strong agreement between predictions from the PINNs and the FEM results. Furthermore, the normalized loading forces predicted from PINNs also agree well with the FEM (\textbf{Fig.2(d)}), demonstrating the strain-stiffening effect in most elastomers with chemical cross-links, which is attributed to the parameter $\lambda_m$ in the AB model. To demonstrate the prediction of stress field from PINNs, we plotted the Cauchy stress contour in the loading direction, $\sigma_{11}$, in \textbf{Fig.2(e)}. We found that the Cauchy stress contours agree well between the PINNs and FEM far away from the hole. While the relative error distanced from the hole remains small ($10^{-2}-10^{-3}$), large errors emerge in the proximity of the hole, with an average value of 0.6. This discrepancy is primarily attributed to a stress concentration around the hole, contributing to a decaying field characterized by a substantial displacement gradient \cite{bower2009applied,anand2020continuum}, especially when the sample is under large deformation. To enhance the accuracy within this localized region, it becomes imperative to employ a multiscale approach capable of accurately representing the local decaying field. For example, combining deep neural operator (DeepONet) \cite{lu2021learning} for local surrogate modeling with global FEM solver has recently shown a promising solution to increase the local simulation accuracy \cite{yin2022interfacing}. However, it is worth noting that our domain integration method effectively captures the long-range effect of the hole on the loading curve (\textbf{Fig.2(d)}). Our tests of different material parameters, hyperelastic models, and sample geometries show considerable accuracy in the stress contour away from the inhomogeneities. Therefore, we believe our PINNs framework demonstrates noteworthy precision in assessing both the displacement field and the loading history, and the localized stress concentration will have negligible effects on parameter identification in the inverse problem, to be shown in the following subsections.

\subsection{Identify constitutive parameters for the prototype problem}

In the last subsection, we demonstrated the proposed PINNs framework can efficiently model the forward problem. In the following subsections, we apply the framework to identify the unknown constitutive parameters, $\boldsymbol{\theta}_{\text{const}}=(\mu,\lambda_{m}$), based on synthetic experimental data. Unlike conventional approaches for inverse problems \cite{groetsch1993inverse, yan2024photoacoustic, yan2021abel, bai2023ponderomotive, yan2021generation}, which formulate as an initial value or boundary value problem, PINNs take a distinctive route by minimizing the loss functions inherent in the constitutive relations. In this subsection, we first employed the PINNs framework to the prototype problem described in \textbf{Section 2.2} (\textbf{Fig.2(a)}). We randomly selected 400 sampling points within the subdomain, $0.2 < |X_1| < 0.8$, to calculate the total loading force at each time step using \textbf{Eq.(9)}. We tuned the weights of each loss to maximize the prediction performance with emphasis on $\mathcal{L}^{\text{inc}}$ and $\mathcal{L}^{\text{Loading}}$.  

\textbf{Fig.3} illustrates the predicted values for $\mu$ and $\lambda_{m}$ for three sets of true values. In the training process, the displacements of 200 randomly selected speckles over 21 time steps serve as additional constraints, except for the green curves. Starting with initial values of $(\mu,\lambda_m)_{\text{initial}} = (4.0,7.0)$, the model with the DIC data demonstrates better convergence in less than $10^6$ epochs, compared with the case without the DIC data (green curves). The convergence speed for the inverse problem is reasonable since the forward model also requires at least 10k training epochs to obtain a reasonably accurate deformation field prediction (\textbf{Fig.2(b)}). Interestingly, we also observed that the PINNs model had quicker convergence when predicting $\mu$ over $\lambda_{m}$, which first optimized the material parameters akin to a NH model. Subsequently, the model progressively learned the strain-hardening behaviors by converging $\lambda_{m}$ to the ground truth. This dynamic evolution in parameter optimization highlights the model's capability to adapt and capture intricate material behaviors as training progresses. In this section, we investigated a relatively simple scenario involving a circular hole in the middle of the sample. In such case, we could derive the theoretical approximations by fitting the loading curving to the 1D uniaxial tension solutions, which yield reasonable predictions with $\mu^{est} \approx 0.94$ and $\lambda_m^{est} \approx 2.97$ when the true values are  1.0 and 3.0, respectively. However, most biological soft materials have much more intricate sample geometries, where theoretical estimations or other inverse techniques cannot readily fit their material parameters. In the subsequent section, we will apply the PINNs framework to a more complex scenario, demonstrating that the loading history data cannot sufficiently identify the parameters. Therefore, integrating DIC data is necessary for parameter inversion with the desired accuracy in handling intricate cases.

\begin{figure}[!ht]
    \centering
    \includegraphics[width=1.0\textwidth]{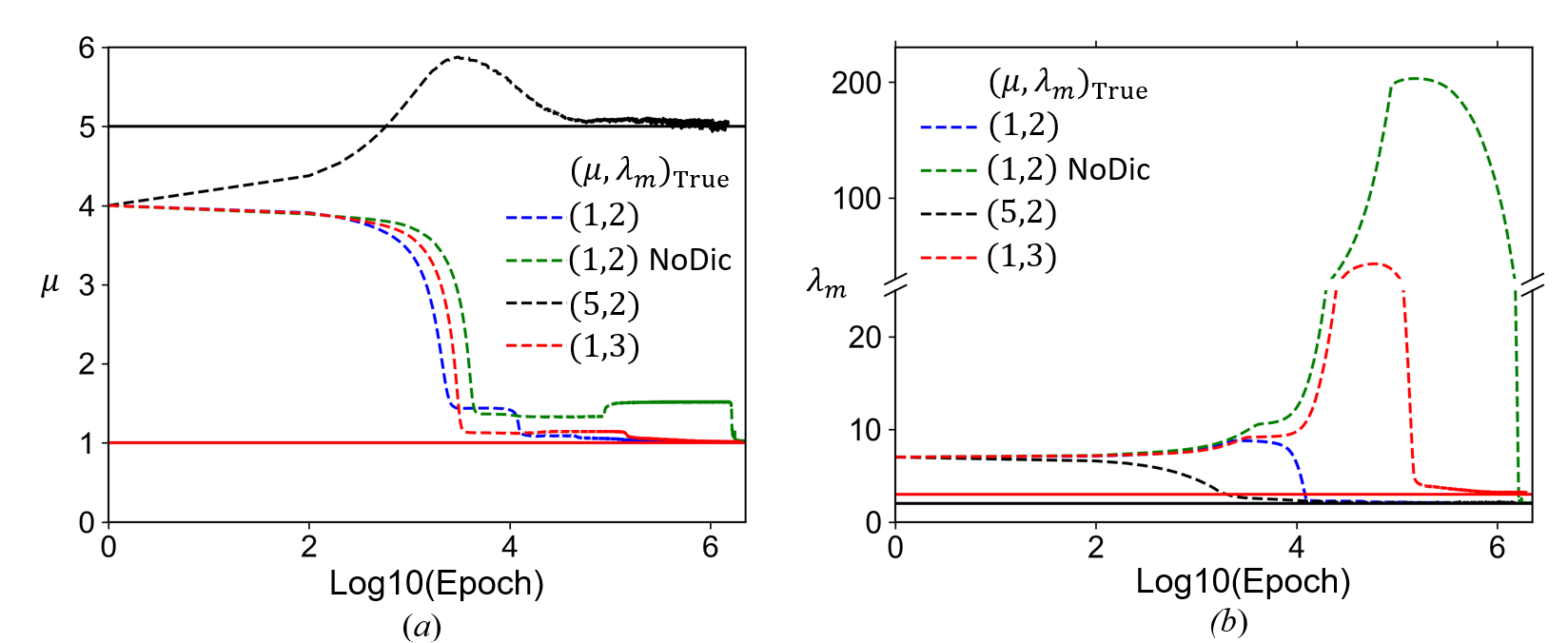}
    \caption{Constitutive parameter identification for a rectangular sample with a central hole under the plane stress condition with different ground truths. (a) Convergence of $\mu$ over 1M training epochs;(b) Convergence of $\lambda_m$ over 1M training epochs. The dashed line is the predictions from PINNs, and the solid line is the ground truth.}
    \label{fig:Figure-3}
\end{figure}

\subsection{Identify constitutive parameters for complex geometries}

In this subsection, we employ the PINNs framework to identify constitutive parameters for a more intricate sample geometry based on both DIC and loading history datasets. The geometry of the sample, involving four internal circular imperfections with the radius of $r=0.15$ situated at coordinates, $(-0.4,0.3)$, $(-0.2,-0.1)$, $(0.1,0.2)$, and $(0.3,-0.3)$, is shown in \textbf{Fig.4(a)}. To increase the complexity of the problem, we set these 4 inhomogeneities as a different material from the medium. Specifically, we chose an incompressible NH material with shear modulus, $\mu_{imp} = 200$, which is unknown in the PINNs calculation. The loading force is evaluated via the domain integral in the region, $0.6<|X_1|<0.8$. We first applied the 1D uniaxial tension approximation to fit the parameters of the medium, and we got $\mu^{est} \approx 1.27$ and $\lambda_m^{est} \approx 2.94$, with the prediction error of 27\%. This large discrepancy emphasizes the challenge posed by complex material and geometry configurations and underscores the need for more comprehensive approaches to accurately determine material parameters. Therefore, we applied our PINNs framework to identify the parameters for this intricate case.

\begin{figure}[!ht]
    \centering
    \includegraphics[width=0.9\textwidth]{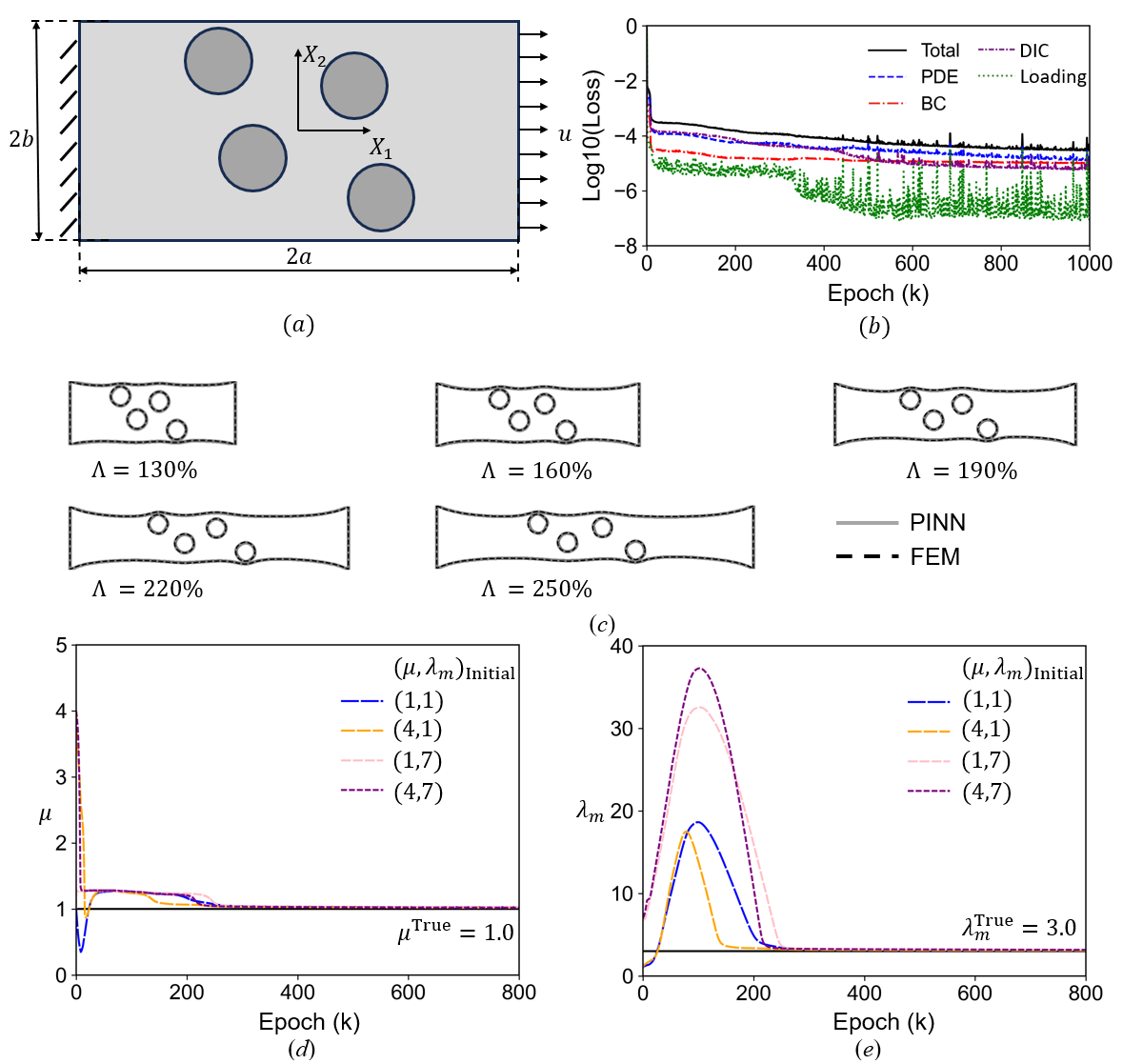}
    \caption{
Constitutive parameter identification for a rectangular sample with four circular inhomogeneities centered at locations: $(-0.4,0.3)$, $(-0.2,-0.1)$, $(0.1,0.2)$, and $(0.3,-0.3)$. (a) Schematics of the sample geometry and dimension; (b) Loss as a function of training epochs. (c) Visual outlines of deformed configurations calculated from PINNs (solid lines) and FEM (dashed lines) under 5 different stretch ratios; (d-e) Constitutive parameter identification of $\mu$ and $\lambda_m$ over 800k training epochs. The dashed line is the predictions from PINNs and the solid line is the ground truth.}
    \label{fig:Figure-4}
\end{figure}

The PDEs were evaluated within the region delineated by light grey shading, as depicted in \textbf{Fig.4(a)}. Furthermore, we incorporated the displacement history of 200 randomly distributed speckles across 21 time steps. Due to the unknown material properties of the four circular inhomogeneities, we utilized the displacement field data obtained from synthetic DIC as BCs for these inhomogeneities. We tuned the weights of loss functions with emphasis on both speckle displacement and the loading history. These optimized weights account for the varying degrees of influence each constraint exerts on the optimization process, effectively guiding the model to converge toward accurate parameter values despite the challenges posed by the complex configuration. We have tested our model with a variety of initial values for the constitutive parameters in a reasonable range and all of them show similar convergence behavior. For illustration, we present the results with the initial values of 
$(\mu,\lambda_m)_{\text{initial}} = (4.0,7.0)$ in \textbf{Figs.4(b)} and \textbf{(c)}. As shown in \textbf{Fig.4(b)}, the losses experience rapid reduction within the first 10k training epochs, followed by a more gradual decline to convergence. In \textbf{Fig.4(c)}, we plotted the displacement outlines predicted by the PINNs framework and the corresponding FEM simulations. Remarkably, despite not explicitly modeling the inherent inhomogeneity, the PINNs adeptly predicted the deformation field based on synthetic DIC data. The predictions for $\mu$ and $\lambda_{m}$ are presented in \textbf{Figs.4(d)}\&\textbf{(e)}. We tested multiple initial values, all of which demonstrated similar convergence performance. Notably, there is a pronounced increase in $\lambda_{m}$ during the initial 180k training epochs, signifying convergence toward the NH limit. As the PDE loss and BC loss gradually diminished, the PINNs framework started to capture the strain-stiffening behavior. The convergence of our model remains stable across various initial values. However, the disparity between initial guesses and the ground truth impacts the rate of convergence. As depicted in \textbf{Fig.3(b)} and \textbf{Fig.4(e)}, the smaller initial value of the $\lambda_{m}$ typically leads to faster convergence. This trend arises because the optimization process tends to overestimate $\lambda_{m}$ in order to minimize errors in the power term, and gradually adjusts its value towards the ground truth. Conversely, if the initial value of $\lambda_{m}$ is set too large, the overestimation becomes more pronounced, resulting in a longer time for convergence.

\subsection{Effects of DIC quality on parameter identification}

Compared with synthetic DIC data generated from FEM, real-world experiments are constrained by limitations in speckle density and measurement resolution. Here, we discuss the effects of speckle density and noise level of synthetic DIC data on the parameter predictions. For each speckle density and noise level, we performed at least 3 calculations with different random seeds. The prediction errors are listed in \textbf{Table 1}, accounting for variations in speckle quantity and noise levels. The convergence rates of several cases are plotted in \textbf{Fig.5}.

\begin{table}[h!]
\centering
\begin{tabular}{ |c|c|c|c|c| }
 \hline
Test Name & Noise level & Number of speckles & $\frac{|\mu - \mu^{\text{True}}|}{\mu^{\text{True}}}$ & $\frac{|\lambda_{m}-\lambda_{m}^{\text{True}}|}{\lambda_{m}^{\text{True}}}$\\
 \hline
PINN-0\%-0   &0      &0  &18.50$\pm$6.25\%&25.26$\pm$10.12\%\\ 
PINN-0\%-40  &0      &40 &2.93$\pm$0.77\%  &4.53$\pm$0.97\%\\
PINN-0\%-200 &0 & 200    &  1.88$\pm$0.95\%   & 3.94$\pm$0.85\% \\
PINN-5\%-40  & $5\%$ & 40   & 3.68$\pm$1.09\%  & 8.16$\pm$2.32\%\\ 
PINN-5\%-200 & $5\%$ & 200 & 2.31$\pm$0.21\%    &  4.12$\pm$0.59\% \\ 
PINN-10\%-40 & $10\%$ &40 &16.04$\pm$9.63\% & Not Converge\\ 
PINN-10\%-200& $10\%$ &200&  3.38$\pm$1.23\%  & 6.99$\pm$2.87\%\\ 
PINN-15\%-40 & $15\%$ &40 & 26.14$\pm$1.46\% &Not Converge\\ 
PINN-15\%-200& $15\%$ &200&7.16$\pm$1.09\% &20.74$\pm$7.48\%\\  
 \hline
\end{tabular}
\caption{PINNs predictions for different speckle density and noise level after 800k epochs.}
\label{table:1}
\end{table}

The results without noise and with different numbers of speckles demonstrate how speckle density affects the parameter prediction accuracy (\textbf{Fig.5(a)}). As we can see, the prediction error for $\mu$ and $\lambda_m$ reaches $18.5\%$ and $25.3\%$ if we do not incorporate synthetic DIC data into training. Remarkably, even with a small number of DIC data points (i.e., 40 speckles), the results show reasonable prediction capabilities with errors less than $5\%$. The results about different noise levels with 200 speckles highlight the impact of noise on parameter identification (\textbf{Fig.5(b)}). As the noise level increases, a marked rise in prediction error is observed. Notably, the prediction of $\lambda_m$ for the cases of 10\% and 15\% noise levels with 40 speckles cannot converge to true values. These results suggest that increasing the speckle density becomes necessary when working with full-field measurement data with high noise levels. Furthermore, the arrangement of speckle positions can influence the training performance. Specifically, increasing the speckle density near the boundaries of the inhomogeneities can provide more accurate displacement information.

\begin{figure}[!ht]
    \centering
    \includegraphics[width=0.9\textwidth]{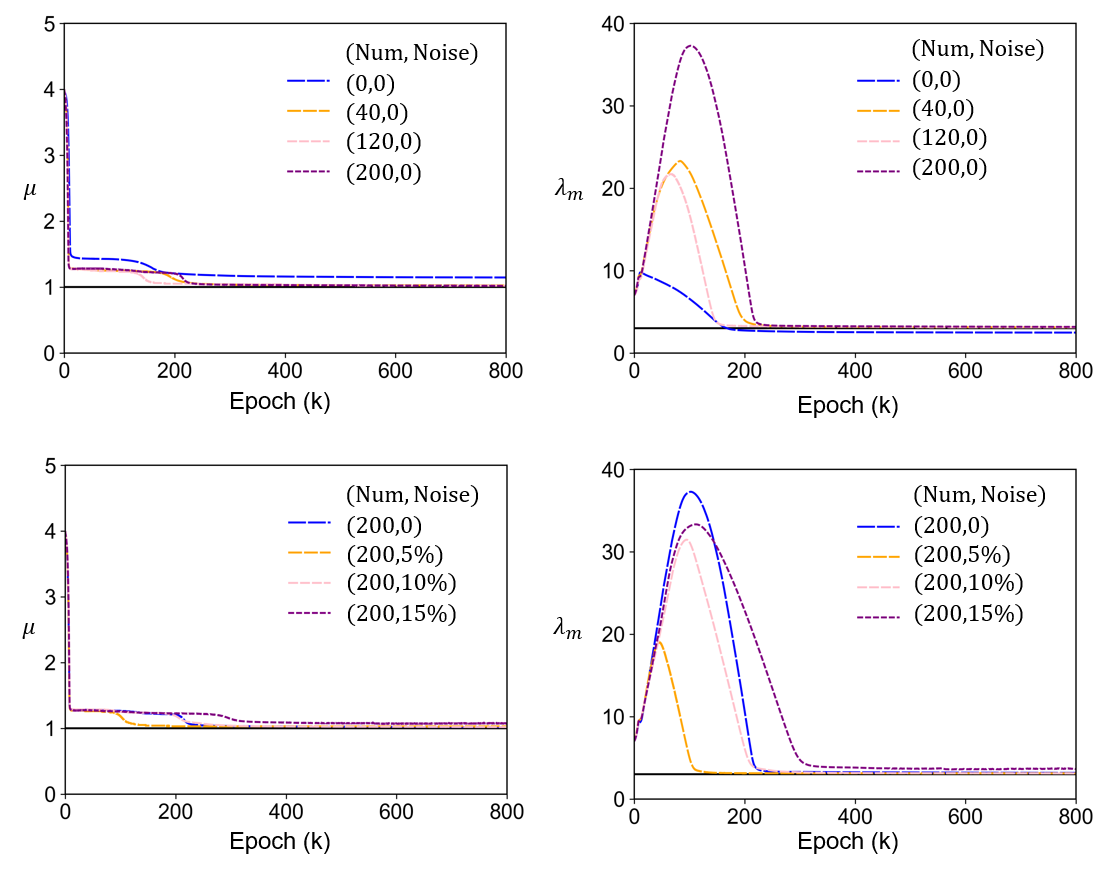}
    \caption{Influence of the speckle density and noise on the predictions of PINNs. (a-i, ii) Speckle density effect. (b-i, ii) Speckle noise level effect.}
    \label{fig:Figure-5}
\end{figure}

Furthermore, we also found that the parameter convergence speed is not directly proportional to the number of speckles (\textbf{Fig.5(a)}). While PINNs model with fewer speckles can sometimes converge quickly, they are more sensititive to noise, especially at higher noise levels, as seen in the cases of PINN-10\%-40 and PINN-15\%-40 (\textbf{Table 1}). This increased sensitivity is partly due to the reduced speckle density, which compromises the information about inhomogeneities. In our calculations, the parameter, $\mu$, consistently shows rapid and stable convergence. This is likely because $\mu$ is closely related to the stress magnitude, which are governed by the loading data. In contrast, training the strain-stiffening parameter, $\lambda_m$, is more time-consuming due to its link to power terms in the constitutive model, which tend to be affected by the noise during training, resulting in a slower convergence. Overall, these results underscore the influence of speckle density and noise levels on the convergence performance of the constitutive parameters.

\section{Conclusions}
In this paper, we proposed a PINN-based inverse framework to identify the constitutive parameters of complex hyperelastic solids under the plane stress condition. As a proof of concept, we focused on the AB model with strain-stiffening effects under large deformation. A distinctive feature of our proposed framework is its emphasis on training the PINNs using multi-modal experimental dataset. By harnessing the comprehensive information encapsulated within the full-field deformation and loading history, our framework has the capability to ensure identifying parameters accurately, even when training with noisy dataset.

We first verified our framework with the forward modeling on a prototype problem and found excellent agreement with FEM results. Then, we applied our framework for parameter inversion with the prototype problem and beyond. Our framework introduced in this paper is particularly suited for cases involving intricate sample geometries that cannot be readily addressed by theoretical estimations or other inverse techniques. Our model pinpointed the necessary parameters while ensuring errors were below 5\%. It is particularly noteworthy that this impressive accuracy was maintained even when the framework was trained with an experimental noise level of 5\% with sufficient full-field measurement density. 

The current PINNs framework can be readily extended to other classes of hyperelastic models, such as Yeoh model \cite{yeoh1993some}, HGO model \cite{holzapfel2000new}, Gent model \cite{gent1996new}, etc. For these constitutive relations emphasizing the strain-stiffening anisotropic behaviors, we anticipate they will show similar behavior to the AB model in that the PINNs converge to an NH limit in the first few epochs and gradually learn the stress-strain curves. Furthermore, the present PINNs framework is versatile and can be extended to accommodate rate-dependent hyperelastic models, including the Bergstrom-Boyce model \cite{bergstrom1998constitutive}. This adaptability is facilitated by the ability to reformulate the PDEs into an incremental form, similar to approaches used in the study of finite deformation plasticity \cite{niu2023modeling}. By incorporating speckle information into the incremental formulation, we anticipate improvements in convergence speed and applicability to more complex sample structures. In addition to the identification of the materials parameters for constitutive models, the current PINNs framework integrating with DIC data and loading history curves can be applied to extract other material properties, such as cohesive zone parameters \cite{jin2022big, wei2022deep}, residual stress in thin films \cite{rao2020determination,chason2016tutorial}, and the flow resistance \cite{song2018modeling}. 

\section*{Appendix: Identifying material parameters for samples with square inhomogeneities}

The proposed PINNs model is versatile for identifying material parameters from samples with diverse geometries. As an example, we examined a case involving square inhomogeneities which are represented as stiff bodies, each with an edge length of 0.2 (\textbf{Fig.A1}). 200 random speckle points were added as BCs with no noise. As shown in \textbf{Fig.A1}, after 80k training epochs, the total loss is less than $10^{-4}$. The prediction has a relative error of $\frac{|\mu - \mu^{\text{True}}|}{\mu^{\text{True}}}=13.79\pm0.66\%$, $\frac{|\lambda_{m}-\lambda_{m}^{\text{True}}|}{\lambda_{m}^{\text{True}}}=2.06\pm1.69\%$. Compared to the samples with circular holes (\textbf{Fig.4}), the stress field near the corners of the square inhomogeneities exhibits singularity, resulting in slower convergence and lower prediction accuracy especially for $\mu$. Despite the complexity of these defects, the current PINNs framework remains stable, providing reasonable predictions. To enhance the accuracy due to the stress concentration, a multiscale approach which can accurately represent the local stress field can be employed in the future work \cite{yin2022interfacing}. 

\renewcommand\thefigure{A1} 
\begin{figure}[!htb]
    \centering
    \includegraphics[width=0.9\textwidth]{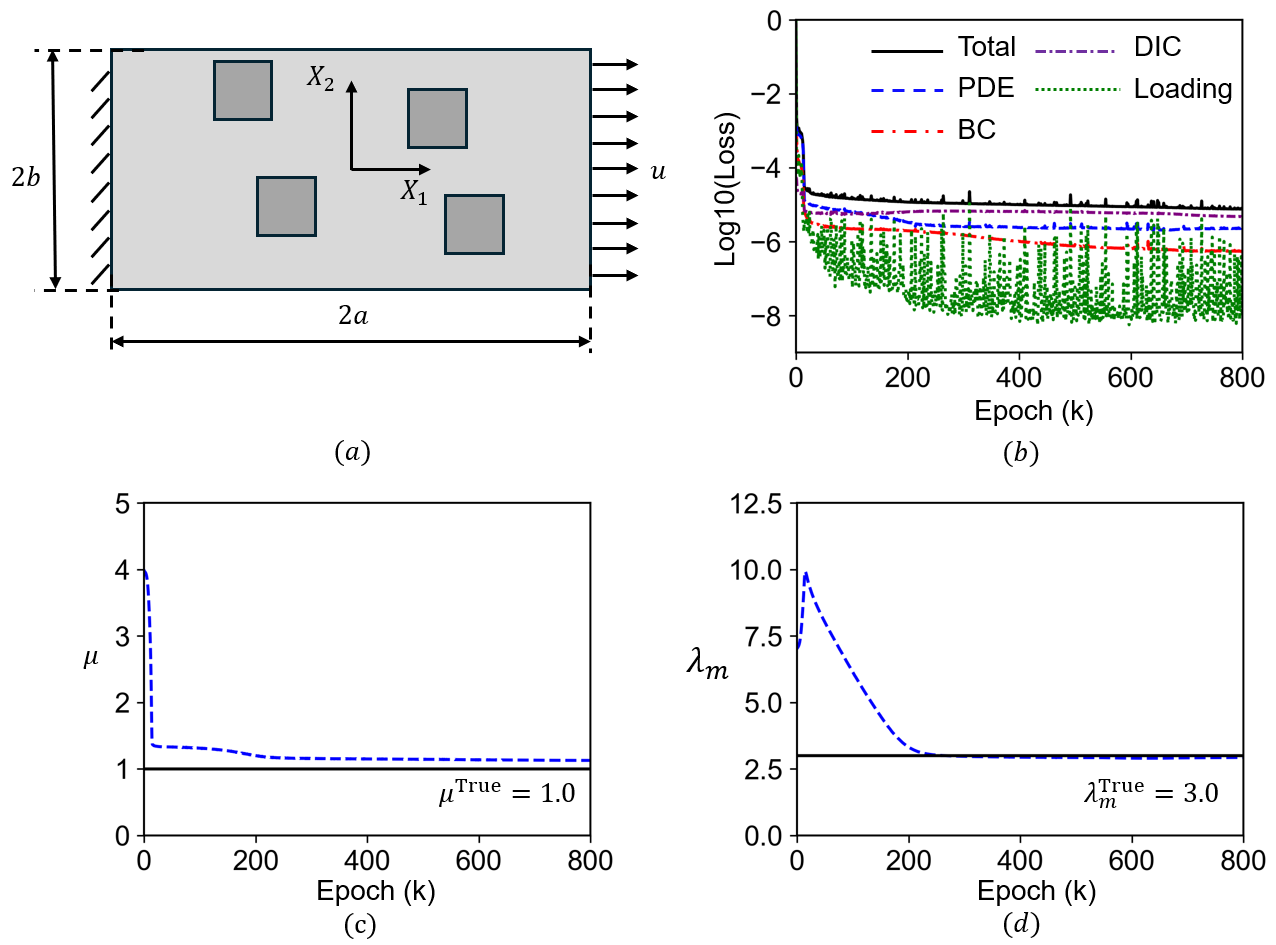}
    \caption{
    Constitutive parameter identification for a rectangular sample with four square inhomogeneities. (a) Schematic of the sample's geometry and dimension: the light grey region represents AB material with the material parameters $
    (\mu,\lambda_{\text{m}})$, while the dark grey region represents incompressible NH material. (b) Loss as a function of training epochs. (c) The prediction of $\mu$ as a function of the training epochs. (d) The prediction of $\lambda_{\text{m}}$ as a function of the training epochs. The dashed line is the predictions from PINNs and the solid line is the ground truth.    
    }
    \label{fig:figure-A}
\end{figure}

\section*{Author Contribution}
H.J. and S.S. conceived the initial idea to perform this research. S.S. and H.J. developed the code. H.J. wrote the initial draft for Section 1 and Section 2. S.S. wrote the initial draft for Section 3 and Section 4. H.J. revised the Section 3 and Section 4. All authors contributed to discussing the results, revising the manuscript, and giving final approval for the publication.

\section*{Data Availability}
The code is publicly accessible on the author's GitHub at:

https://github.com/ssong26/DeepModulus.git.

\section*{Acknowledgement}
Helpful discussions with Dr. Enrui Zhang, and Mr. Xincheng Wang are acknowledged. This research was supported in part through the computational resources provided for the Quest high performance computing facility at Northwestern University which is jointly supported by the Office of the Provost, the Office for Research, and Northwestern University Information Technology.

\section*{Declaration of generative AI and AI-assisted technologies in the writing process}
During the preparation of this work the authors used ChatGPT4.0 for language polishing. After using this tool, the authors reviewed and edited the content as needed and take full responsibility for the content of the publication.

\bibliographystyle{elsarticle-num-names}
\bibliography{reference.bib}

\end{document}